\documentclass[usenatbib,referee]{mn2e}
\usepackage{graphicx}
\usepackage{epsfig}
\usepackage{color}
\usepackage{enumerate}
\usepackage{hyperref}

\title[The formation of single NSs] 
{The formation of single neutron-stars from double white-dwarf mergers via accretion-induced collapse}
\author[D. Liu, B. Wang]
{D. Liu$^{\rm 1,2,3,4}$\thanks{E-mail:liudongdong@ynao.ac.cn}, B. Wang$^{\rm 1,2,3,4}$\thanks{E-mail:wangbo@ynao.ac.cn}\\
$^1$Yunnan Observatories, Chinese Academy of Sciences, Kunming 650216, China\\
$^2$Key Laboratory for the Structure and Evolution of Celestial Objects, Chinese Academy of Sciences, Kunming 650216, China\\
$^3$University of Chinese Academy of Sciences, Beijing 100049, China\\
$^4$Center for Astronomical Mega-Science, Chinese Academy of Sciences, Beijing, 100012, China}

\begin{document}
\date{}
\pagerange{\pageref{firstpage}--\pageref{lastpage}} \pubyear{2020}
\maketitle

\label{firstpage}

\begin{abstract}\label{0. abstract}
The merging of double white-dwarfs (WDs) may produce the events of accretion-induced collapse (AIC) and form single neutron stars (NSs).
Meanwhile, it is also notable that the recently proposed WD$+$He subgiant scenario has a significant contribution to the production of massive double WDs, in which the primary WD grows in mass by accreting He-rich material from a He subgiant companion.
In this work, we aim to study the binary population synthesis (BPS) properties of AIC events from the double WD mergers by considering the classical scenarios and also the contribution of the WD$+$He subgiant scenario to the formation of double WDs.
Firstly, we provided a dense and large model grid of WD$+$He star systems for producing AIC events through the double WD merger scenario. Secondly, we performed several sets of BPS calculations to obtain the rates and single NS number in our Galaxy.
We found that the rates of AIC events from the double WD mergers in the Galaxy are in the range of 1.4$-$$8.9\times10^{\rm -3}\,\rm yr^{\rm -1}$ for all ONe/CO WD$+$ONe/CO WD mergers, and in the range of 0.3$-$$3.8\times10^{\rm -3}\,\rm yr^{\rm -1}$ when double CO WD mergers are not considered. We also found that the number of single NSs from AIC events in our Galaxy may range from $0.382\times10^{\rm 7}$ to $1.072\times10^{\rm 8}$. 
The chirp mass of double WDs for producing AIC events distribute in the range of 0.55$-$$1.25\,\rm M_{\odot}$. We estimated that more than half of double WDs for producing AIC events are capable to be observed by the future space-based gravitational wave detectors.
\end{abstract}


\section{Introduction} \label{1. Introduction}
Double white-dwarfs (WDs) are one of the possible outcomes of low-/intermediate-mass binaries. The double WDs rotate each other, and may eventually merge due to the gravitational wave radiation.
The mergers of double WDs are relevant to many peculiar events, such as (1) hot subdwarfs from the mergers of
double He WDs (e.g. Han et al. 2003; Zhang \& Jeffery 2012), (2) type Ia supernovae from the mergers of double CO WDs (e.g. Iben \& Tutukov 1984; Webbink 1984; Nelemans et al. 2001; Liu et al. 2018a) or the mergers between CO WDs and He-rich WDs (e.g. Dan et al. 2012; Crocker et al. 2017; Liu et al. 2017), (3) extreme He stars and R Coronae Borealis stars from the mergers of CO WDs with He WDs (e.g. Webbink 1984; Iben \& Tutukov 1985; Zhang et al. 2014), (4) accretion-induced collapse (AIC) events from the mergers involving double ONe WDs, ONe WD$+$CO WD and double CO WDs (e.g. Nomoto \& Iben 1985; Saio \& Nomoto 1985; Ruiter et al. 2019), etc. The merging process of double WDs may also relate to some high-energy phenomena, such as Gamma-ray burst, high-energy neutrino emission, etc (e.g. Lyutikov \& Tonoon 2017; Xiao et al. 2016). Meanwhile, the merging of double WDs, including double WD systems themselves, are verified gravitational wave sources (e.g. Kilic et al. 2014), and they may be observable by the space-based GW detectors like LISA (e.g. Ruiter et al. 2010; Kremer et al. 2017).

The merging of double WDs can form single neutron stars (NSs) via the AIC process (e.g. Ruiter et al. 2019)\footnote{Note that NSs can be formed through the core-collapse supernovae of massive stars, the electron capture supernovae of intermediate-mass stars and the AIC events of massive WDs (e.g. van den Heuvel 2009).}. During the merging process, the secondary WD (i.e. the less-massive WD) may be torn apart, and form a hot envelope or a thick disc, or even both on the surface of the primary WD (i.e. the massive WD; e.g. Kashyap et al. 2015). For the ONe WD$+$ONe/CO WD mergers, the primary ONe WD would collapse into an NS caused by the electron-capture reactions of Mg and Ne when the mass of the merging remnant exceeds or approach to the Chandrasekhar mass limit, i.e. the double-degenerate (DD) model of AIC events (e.g. Nomoto \& Iben 1985; Wu \& Wang 2018a), although it has been argued that the mergers of ONe WD$+$CO WD systems may also produce the faint and rapid type Ia supernovae via the failed detonation process (e.g. Kashyap et al. 2018).
Meanwhile, an ONe WD can also grow in mass by accreting material from a main-sequence (MS) star, or red-giant (RG) star, or a He star in the single-degenerate (SD) model. When the ONe WD grows in mass close to the Chandrasekhar mass limit, it may collapse to a NS (e.g. Canal et al. 1990; Yungelson \& Livio 1998; Tauris et al. 2013; Ablimit \& Li 2015; Brooks et al. 2017; Wang 2018a). Wang, Podsiadlowski \& Han (2017) suggested that CO WD$+$He star systems could also produce AIC events for the case of off-center C ignition when the WD grows in mass close to the Chandrasekhar mass limit.

In addition, the merging of double CO WDs may also produce single NSs through the AIC process (e.g. Timmes, Woosley \& Taam 1994; Podsiadlowski et al. 2004; Shen et al. 2012; Schwab, Quataert \& Bildsten 2015; Wu, Wang \& Liu 2019), although it is a relatively promising scenario for producing type Ia supernovae (e.g. Iben \& Tutukov 1984; Webbink 1984; Nelemans et al. 2001; Ruiter et al. 2009; Toonen, Nelemans \& Portegies 2012; Liu et al. 2018a; Wang 2018b). The outcomes of double CO WD mergers are still under hot debate. Pakmor et al. (2010) recently suggested that the mergers of double CO WDs can produce type Ia supernovae only when the process is violent under certain conditions, i.e. the violent merger scenario for the progenitors of type Ia supernovae (see also Pakmor et al. 2011, 2012; Ruiter et al. 2013). According to these results, it can be simply assumed that double CO WD mergers can produce single NSs via the AIC process when the merging process is not violent. Note that the double-degenerate core of a planetary nebula Henize 2-428 is a candidate for the progenitor of type Ia supernova based on the violent merger scenario (Santander-Garc\'ia et al. 2015), while KPD 1930$+$2752 is a massive WD$+$sdB star system (Maxted et al. 2000; Geier et al. 2007), which may evolve to double WDs and merge to form an AIC event.

Recently, an evolutionary scenario named the WD$+$He subgiant scenario has been proposed as an important path for the formation of double massive WDs (Ruiter et al. 2013; Liu et al. 2018a). In this scenario, the He star transfers He-rich matter onto the surface of the primary WD when the it fills its Roche-lobe, leading to the mass-growth of the WD. The binary eventually evolves to a double WD system when the He subgiant exhausts its He-shell.
Liu et al. (2018a) found that the predicted delay-time distributions of type Ia supernovae from the classical DD model can match better with the observations after considering this evolutionary scenario, especially for the observed type Ia supernovae in old epoches and early epoches that cannot be explained by previous studies.
The present work will provide the parameter space of WD$+$He star systems that can evolve to double WDs (including double ONe WDs, ONe WD$+$CO WD and double CO WDs) and then produce single NSs via the AIC process when they merge.

This paper is organized as follows. In Sect.\,2, we provide the parameter space of WD$+$He star systems for producing double WDs that can merge and produce AIC events. The binary population synthesis (BPS) calculations are shown in Sect.\,3. We present a discussion in Sect.\,4, and finally a summary in Sect.\,5.

\section{binary evolution simulations}\label{binary evolution calculations}
\subsection{Binary evolution methods}
In order to obtain the contribution of double WD mergers from the WD$+$He subgiant scenario for the production of AIC events, we employ the Eggleton stellar evolution code and evolved a large number of WD$+$He star systems to the formation of double WDs (see Eggleton 1973; Han, Podsiadlowski \& Eggleton 1994; Pols et al. 1995, 1998; Eggleton \& Kiseleva-Eggleton 2002). We have evolved a large number of WD$+$He star systems that can form double WDs in our previous works (Liu et al. 2018a). In the present work, we use these binary evolution results, and evolve a large number of supplementary WD$+$He star systems. The assumptions in the stellar evolution code are similar to our previous work in Liu et al. (2018a).

The criteria for the formation of AIC events from double WD mergers in the present work are described as follows: (1) The double WDs should be double ONe WDs, ONe WD$+$CO WD or double CO WDs. (2) The total masses of the double WDs should be larger than the Chandrasekhar mass limit, which is adopted to be $1.378\,\rm M_{\odot}$. (3) For the case of double CO WD mergers, the double WDs should satisfy one of the following two conditions: (i) the mass ratio, $q_{\rm CO}=M_{\rm WD2}/M_{\rm WD1}$, should be less than 0.8, where $M_{\rm WD1}$ and $M_{\rm WD2}$ are the mass of the massive WD and less massive WD, respectively; (ii) the masses of the massive WDs should be less than $0.8\,\rm M_{\odot}$. This is because the merger of double CO WDs will be violent when the double CO WDs do not satisfy both of these two conditions simultaneously, which may lead to the formation of type Ia supernovae (e.g. Pakmor et al. 2010, 2011; Sim et al. 2010). However, the outcomes of double CO WD mergers are still quite uncertain. In Sect.\,3, we will also consider the case without double CO WD mergers.

In the present work, we provided a dense and large model grid of WD$+$He star systems for the formation of AIC events via the double WD merger scenario. In our calculations, the initial masses of the primary ONe WDs ($M_{\rm WD}^{\rm i}$) is set to be from $1.0$ to $1.3\,\rm M_{\odot}$, and the initial masses of the primary CO WDs range from $0.5$ to $1.2\,\rm M_{\odot}$. In order to produce AIC events via the DD model, the initial masses of the He stars ($M_{\rm 2}^{\rm i}$) varies from $0.3$ to $2.6\,\rm M_{\odot}$, and the initial orbital periods of the WD$+$He star systems ($P^{\rm i}$) are in the range of $0.04-0.50\,\rm d$.

\subsection{Binary evolution results}
Fig.\,1 shows a typical example for the evolution of an ONe WD$+$He star system with ($M_{\rm WD}^{\rm i}$, $M_{\rm 2}^{\rm i}$, $\log\,P^{\rm i})=(1.2, 0.9, -0.69)$ that can form an ONe WD$+$CO WD system and then produce an AIC event, where the mass of WD and He star are in units of $\rm M_{\odot}$ and the orbital period is in units of days.
Firstly, the He MS star evolves to its subgiant stage and expands quickly. At this stage, the He star will fill its Roche-lobe and transfer He-rich material onto the surface of the ONe WD. In this case, the mass-transfer rate is in the order of $10^{\rm -7}\,\rm M_{\odot}yr^{\rm -1}$ and decrease with time. The He shell flashes weakly and part of the He-rich material burn to O and Ne, leading to the mass-accumulation onto the surface of the ONe WD. At the end of the mass-transfer process, the He shell turns to flash so strong that almost no He-rich material accumulated. After about $4.33 \times 10^5\,\rm yr$, the He shell in the He star is exhausted and turns to be a CO WD. At this moment, a system consisted of a $1.3055\,\rm M_{\odot}$ ONe WD and a $0.7614\,\rm M_{\odot}$ CO WD is formed. After about $2.3 \times 10^9\,\rm yr$, the orbital period is $\log(P^{\rm f}/\rm day)=-0.8658$ when our simulation terminated. Subsequently, the double WDs continues to cool down and will eventually merge due to the gravitational wave radiation in about $2.4\times 10^9\,\rm yr$, resulting in the production of an AIC event.

\begin{figure*}
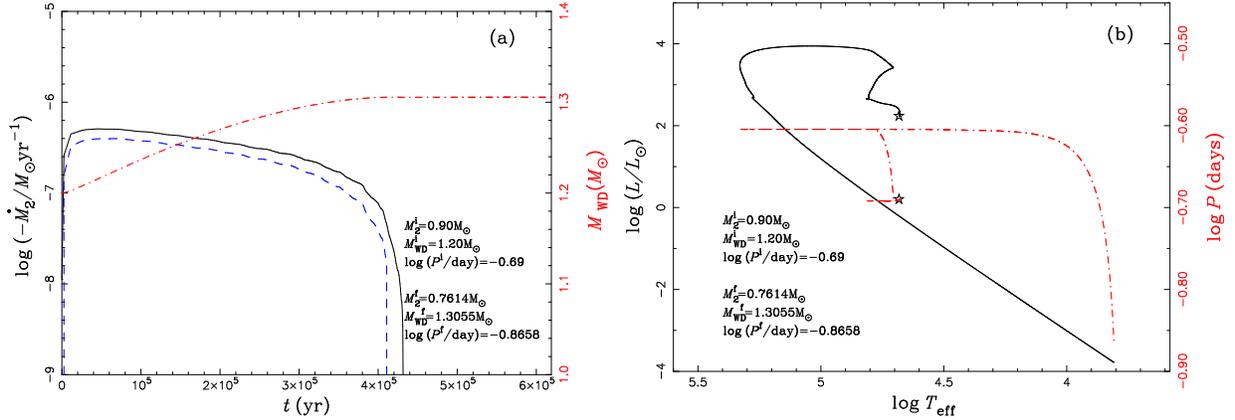

\centerline{\epsfig{file=f1a.ps,angle=270,width=8cm}\ \ \epsfig{file=f1b.ps,angle=270,width=8cm}} \caption{An
  example for the evolution of a ONe WD$+$He star system that can
  evolve to an ONe WD$+$CO WD system and then form an AIC event. Panel (a) shows
  the evolution of the mass transfer rate (solid line), the WD
  mass-growth rate (dashed line) and the WD mass (dash-dotted line)
  changing with time. Panel (b) shows the
  luminosity of the mass donor (the solid line) and the binary orbital
  period (the dash-dotted line) as a function of effective temperature. The open stars in the right panel indicate the position where the simulation starts.}
\end{figure*}

Fig.\,2 presents the final outcomes of ONe WD$+$He star systems in the $\log P^{\rm i}-M^{\rm i}_2$ plane. Those ONe WD$+$He star systems will evolve to ONe WD$+$CO WD systems and then merge to produce AIC events (filled circles in the red solid grids), or evolve to double ONe WD systems and then form AIC events (asterisks in the blue dotted grids). Other binaries cannot evolve to AIC events via the DD model.
Binaries represented by crosses will evolve to double WDs with total masses less than the Chandrasekhar limit; the mergers of these double WDs might form a single massive WD like RE J0317$-$853 (e.g. Barstow et al. 1995; K\"ulebi et al. 2010; Tout et al. 2008; Maoz, Mannucci \& Nelemans 2014; Cheng et al. 2019, 2020).
For the binaries denoted by triangles, the delay times from the primordial binaries to the formation of AIC events are larger than the Hubble time.
The ONe WDs in the binaries indicated by squares will increase their masses to the Chandrasekhar limit by accreting He-rich material from the He companions, i.e. the SD model for producing AIC events (see Liu et al. 2018b; Wang 2018a). Binaries marked by open circles will undergo a dynamical unstable mass transfer process when the He subgiants fill their Roche-lobe, leading to common envelope (CE) phases. After the CE ejection, these systems may also evolve to double WDs and produce AIC events, which will be discussed as part of the CE ejection scenarios for the formation of double WDs in Sect.\,3.
For the binaries represented by pluses, their outcomes may be quite complex. Their orbital separation are quite close, and thus the mass transfer rate is relatively low. The accreted He will form thick He layers on the surface of the WDs. The He layers may trigger detonations when they are thick enough (e.g. Nomoto 1982; Woosley, Taam \& Weaver 1986). Marquardt et al. (2015) argued that the detonations of ONe WDs triggered by the first He detonation may lead to the formation of type Ia supernovae.


\begin{figure*}
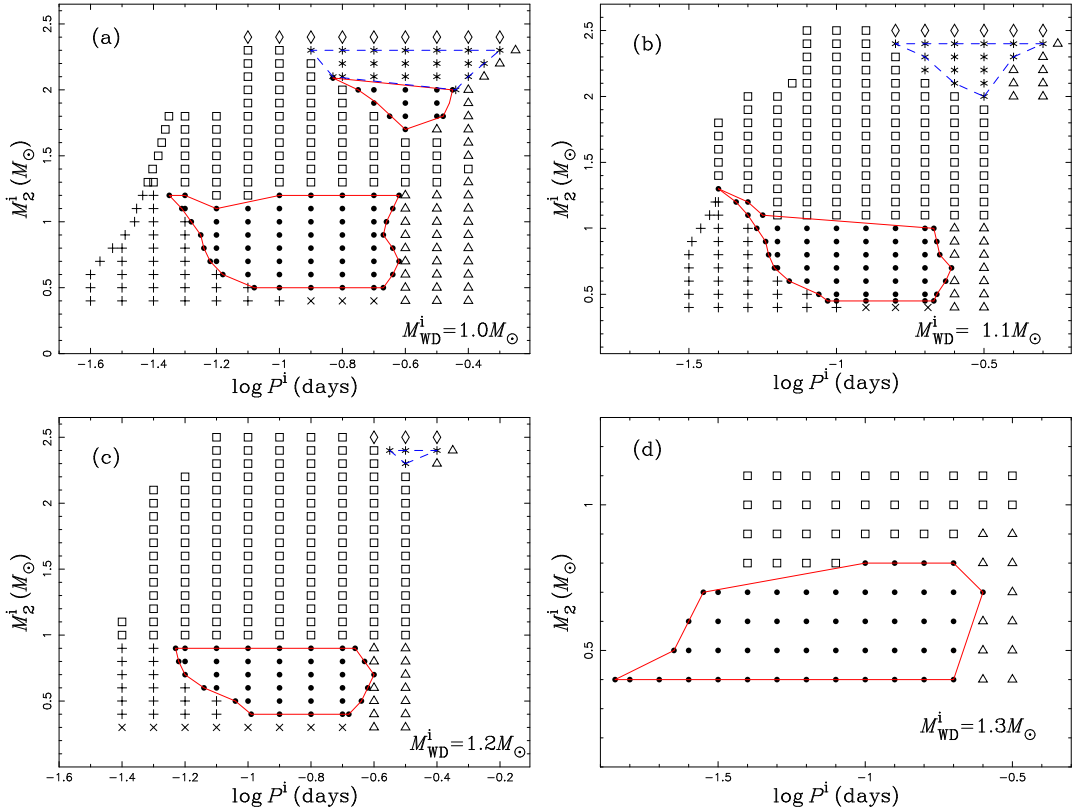

\begin{tabular}{@{}cc@{}}
\centerline{\epsfig{file=f2a.ps,angle=270,width=7.cm}\ \ \epsfig{file=f2b.ps,angle=270,width=7.cm}} \\
\centerline{\epsfig{file=f2c.ps,angle=270,width=7.cm}\ \ \epsfig{file=f2d.ps,angle=270,width=7.cm}}
\end{tabular}
\caption{The grids of ONe WD$+$He star systems for producing AIC events in the initial orbital period$-$initial He star mass ($\log P^{\rm i}-M^{\rm i}_2$) plane. Every panel presents the outcomes of a particular initial WD mass. The filled circles in the red solid grids represent ONe WD$+$He star systems that would evolve to ONe WD$+$CO WD systems, while asterisks in the blue dotted grids stand for binary that would evolve to double ONe WD systems. Other binaries cannot produce AIC events via the DD model; they may form double WDs less massive than the Chandrasekhar limit (crosses), or have delay times larger than the Hubble time (triangles), or produce AIC events through the SD model (squares), or undergo a dynamical unstable mass transfer process (open circles), or experience He layer detonation which may be quite complex (pluses).}
\end{figure*}

\begin{figure*}
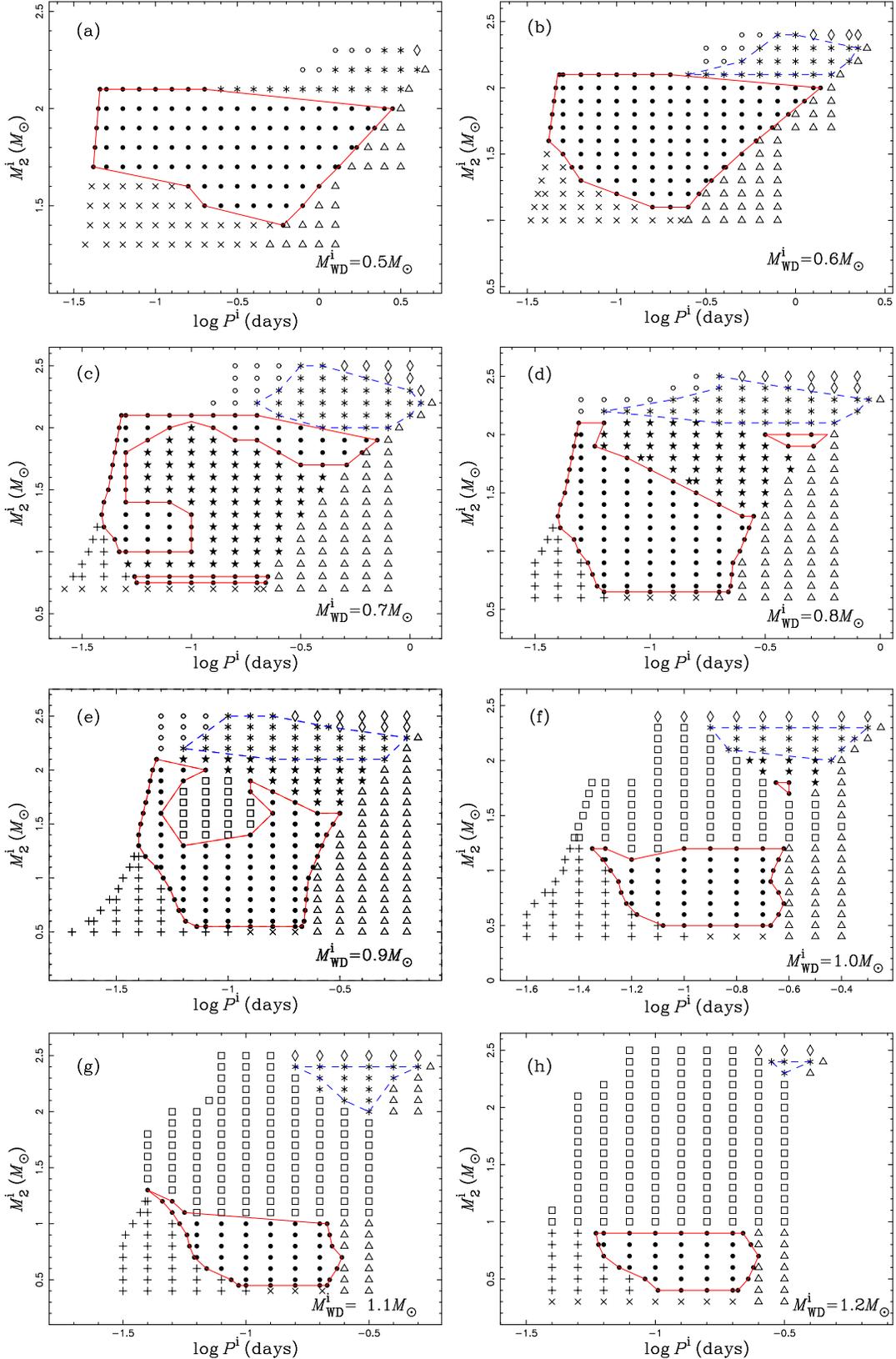

\begin{tabular}{@{}cc@{}}
\centerline{\epsfig{file=f3a.ps,angle=270,width=7.cm}\ \ \epsfig{file=f3b.ps,angle=270,width=7.cm}} \\
\centerline{\epsfig{file=f3c.ps,angle=270,width=7.cm}\ \ \epsfig{file=f3d.ps,angle=270,width=7.cm}} \\ \centerline{\epsfig{file=f3e.ps,angle=270,width=7.cm}\ \ \epsfig{file=f3f.ps,angle=270,width=7.cm}} \\ \centerline{\epsfig{file=f3g.ps,angle=270,width=7.cm}\ \ \epsfig{file=f3h.ps,angle=270,width=7.cm}}
\end{tabular}
\caption{The grids of CO WD$+$He star systems for producing AIC events in the $\log P^{\rm i}-M^{\rm i}_2$ plane. Every panel presents the outcomes of a particular initial WD mass. The filled circles in the red solid grids represent CO WD$+$He star systems that would evolve to double CO WD systems, while asterisks in the blue dotted grids stand for binary that would evolve to CO WD$+$ONe WD systems. Other binaries cannot produce AIC events via the DD model; they may form double WDs less massive than the Chandrasekhar limit (crosses), or have delay times larger than the Hubble time (triangles), or produce AIC events via the SD model (squares), or undergo dynamical unstable mass transfer process (open circles), or produce type Ia supernovae via the double-detonation model (pluses) or the violent merger model (five-pointed stars). Part of these data originate from Liu et al. (2018a).}
\end{figure*}

Fig.\,3 shows the final outcomes of CO WD$+$He star systems in the $\log P^{\rm i}-M^{\rm i}_2$ plane. The filled circles in the red solid grids represent CO WD$+$He star systems that will evolve to double CO WDs and then merge to form AIC events, while the asterisks in the blue dotted grids stand for those will evolve to CO WD$+$ONe WD systems. The filled stars, squares and pluses represent CO WD$+$He star systems that may respectively produce type Ia supernovae through the violent merger model (see Liu et al. 2018a), the SD model (see Wang et al. 2009) and the double-detonation model (see Wang, Justham \& Han 2013).

It may hardly produce CO WD$+$ONe WD systems from the CO WD$+$He star systems.
This is because the masses of primordial stars for CO WDs are typically less than that for ONe WDs. More massive stars should evolve earlier than less massive ones. However, for the cases here, the primordial stars that will form CO WDs should evolve earlier than that will form ONe WDs. There is one scenario to form CO WD$+$He star systems that can evolve to CO WD$+$ONe WD systems, in which the primordial primary star fills its Roche-lobe and transfers a great deal of material to the primordial secondary, resulting in a mass reversal of the binary, i.e. forming an algol binary. In this scenario, the primordial primary become less massive and can evolve to a CO WD, and the primordial secondary is more massive and could evolve to an ONe WD. This scenario will be included as an alternative formation scenario for ONe WD$+$CO WD systems in Sect.\,3.

It is notable that the primary ONe/CO WDs can grow in mass up to $0.48\,\rm M_{\odot}$ during the mass accretion process. Thus, the WD$+$He subgiant scenario is very important for the formation of double massive WDs (see also Ruiter et al. 2013; Liu et al. 2018a). These double WDs originating from the WD$+$He subgiant scenario have He-rich atmosphere, i.e. those binaries should be double DB/DO WDs. Recently, Genest-Beaulieu \& Bergeron (2019) presented some clear
evidence for 55 unresolved double DB WDs in a photometric and spectroscopic investigation of the DB WD population using SDSS and Gaia data. The discovery of so many double DB WDs is an evidence for the WD$+$He subgiant scenario for the formation of double massive WDs.

\section{Binary population synthesis}\label{Binary population synthesis}
By employing the Hurley rapid binary evolution code, we conduct a series of Monte Carlo simulations to calculate the rate and delay time distribution of AIC events (Hurley, Tout \& Pols 2002). If the primordial binaries can directly evolve to CO/ONe WD$+$CO/ONe WD systems and satisfy the criteria presented in Sect.\,2.1, we assume that AIC events will occur when these double WDs merge (i.e. the CE ejection scenario; see Liu et al. 2018a). Meanwhile, if the primordial binaries can evolve to CO/ONe WD$+$He star systems that are located in the parameter space shown in Figs\,2$-$3, we also assume that AIC events will occur (i.e. the WD$+$He subgiant scenario or the stable Roche-lobe overflow scenario).
The basic assumptions and input initial parameters are similar to our previous studies (e.g. Liu et al. 2018a).

The CE ejection process is crucial for the BPS simulations, and is still under highly debate (e.g. Ivanova et al. 2013). In the present work, we adopt the standard energy prescription from Webbink (1984) to calculate the CE ejection process. There are two uncertain parameters in this prescription, i.e. the CE ejection efficiency ($\alpha_{\rm CE}$) and the stellar structure parameter ($\lambda$). Here, we set $\alpha_{\rm CE}\lambda=0.5, 1.0$ and 1.5 for comparison.
In Tab.\,1, we present the Galactic rate, NS number and delay time distribution of AIC events from different merger scenarios with different CE ejection parameters.

\begin{table*}
\begin{center}
 \caption{The Galactic rate, Galactic NS number and delay time distribution of AIC events from different double WD merger scenarios with different CE ejection parameters.} 
   \begin{tabular}{ccccccccc}
\hline \hline
 Set & $\rm Channels$ & $\alpha_{\rm CE}\lambda$ & $\rm Galactic\,\,rates$ & $\rm Galactic\,\,NS\,\,Number$ & $\rm DTDs$\\
 & & & $(10^{\rm -3}{\rm yr}^{\rm -1})$ & $(10^{\rm 7})$ & $({\rm Myr})$\\
\hline
$1$ & $\rm The\,\,double\,\,ONe\,\,WD\,\,mergers$ & $0.5$ & $0.051$ & $0.061$ & $>55$\\
$2$ & $\rm The\,\,double\,\,ONe\,\,WD\,\,mergers$ & $1.0$ & $0.216$ & $0.259$ & $>45$\\
$3$ & $\rm The\,\,double\,\,ONe\,\,WD\,\,mergers$ & $1.5$ & $0.285$ & $0.342$ & $>56$\\
$4$ & $\rm The\,\,ONe\,\,WD$+$\rm CO\,\,WD\,\,mergers$ & $0.5$ & $0.268$ & $0.321$ & $>55$\\
$5$ & $\rm The\,\,ONe\,\,WD$+$\rm CO\,\,WD\,\,mergers$ & $1.0$ & $1.773$ & $2.128$ & $>55$\\
$6$ & $\rm The\,\,ONe\,\,WD$+$\rm CO\,\,WD\,\,mergers$ & $1.5$ & $3.486$ & $4.184$ & $>55$\\
$7$ & $\rm The\,\,double\,\,CO\,\,WD\,\,mergers$ & $0.5$ & $1.129$ & $1.356$ & $>90$\\
$8$ & $\rm The\,\,double\,\,CO\,\,WD\,\,mergers$ & $1.0$ & $3.080$ & $3.696$ & $>110$\\
$9$ & $\rm The\,\,double\,\,CO\,\,WD\,\,mergers$ & $1.5$ & $5.160$ & $6.194$ & $>110$\\
\hline \label{1}
\end{tabular}
\end{center}
\end{table*}

\begin{figure*}
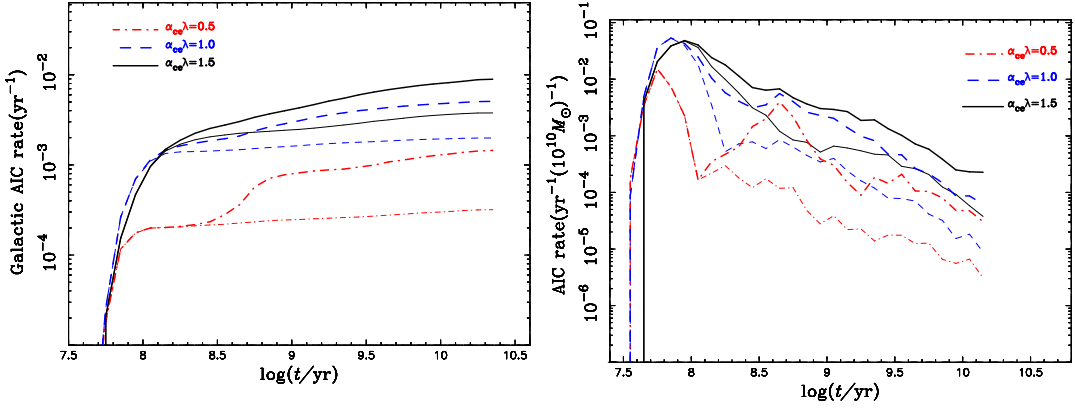

\centerline{\epsfig{file=f4a.ps,angle=270,width=7.cm}\ \ \epsfig{file=f4b.ps,angle=270,width=7.cm}}
 \caption{Evolution of Galactic AIC rates as a function of time based on the DD model by assuming a constant star formation rate of $5\,\rm M_{\odot} yr^{\rm -1}$ (left panel), and the delay time distribution of AIC events by assuming a single starburst with a total mass of $10^{\rm 10}\,\rm M_{\odot}$. The thick lines represent the case producing AIC events from all ONe/CO WD$+$ONe/CO WD mergers, and the thin lines show the case without double CO WD mergers.} 
\end{figure*}

Fig.\,4 shows the evolution of Galactic rates of AIC events in the left panel, and the delay time distribution of AIC events in the right panel with different values of $\alpha_{\rm CE}\lambda$. From the left panel, we can see that the Galactic rate of AIC events are in the range of $\sim$1.4$-$$8.9\times10^{\rm -3}\,\rm yr^{\rm -1}$ for all ONe/CO WD$+$ONe/CO WD mergers, and range from $\sim$0.3$-$$3.8\times10^{\rm -3}\,\rm yr^{\rm -1}$ without considering double CO WD mergers. That means the Galactic rates of AIC events from double WD mergers is about 10\% to 2 times of the observed SN Ia rate.
Wang (2018a) also present that the Galactic rates of AIC events from the SD model are in the range of $\sim$0.3$-$$0.9\times10^{\rm -3}\,\rm yr^{\rm -1}$. Hence, the total Galactic rates of AIC events from both the SD model and the DD model are $\sim$1.8$-$$9.8\times10^{\rm -3}\,\rm yr^{\rm -1}$, or in the range of $\sim$0.6$-$$4.7\times10^{\rm -3}\,\rm yr^{\rm -1}$ when double CO WD merger is not considered. This quite high rate indicates that direct detection of AIC events could be expected. We will discuss the detections of AIC events in Sect.\,4.2.

The delay time distribution of AIC events are defined as the time interval from the formation of primordial binaries to the occurrence of AIC events. From the right panel of Fig.\,4, we can see that the delay times of AIC events from the DD model are from $45\,\rm Myr$ to the Hubble time. The minimum delay time here is mainly determined by the MS lifetime of the most massive stars that can form ONe WDs, as discussed in previous studies (e.g. Doherty et al. 2017; Wang 2018a).

\begin{figure}
\begin{center}
\epsfig{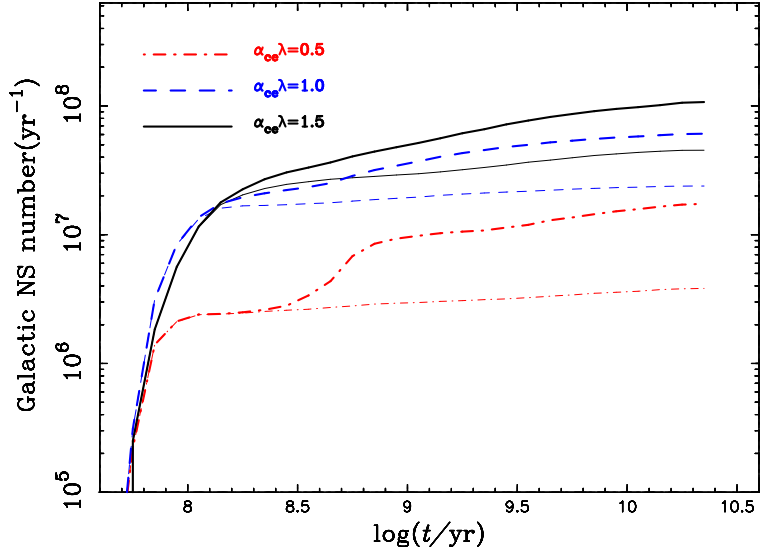}
 \caption{Similar to the left panel of Fig.\,4, but for the evolution of Galactic number of single NSs originating from the AIC events.}
  \end{center}
\end{figure}

The merging of double WDs will produce single NSs after the AIC process.\footnote{Note that ONe WD$+$RG systems may evolve to NS$+$RG systems with large orbital periods. Such wide NS binaries may be disrupted by stellar encounters in globular clusters (e.g. Verbunt \& Freire 2014), which may also lead to the formation of single NSs in globular clusters (e.g. Tauris et al. 2013).} Fig.\,5 shows that the single NS number originating from double WD mergers ranges from $\sim$$1.738\times10^{\rm 7}$ to $1.072\times10^{\rm 8}$ for all ONe/CO WD$+$ONe/CO WD mergers, and are in the range of $\sim$0.382$-$$4.526\times10^{\rm 7}$ without considering double CO WD mergers. According to the results of Wang (2018a), we obtained that the galactic number of NSs from the SD model of AIC events may range from $\sim$$4.054\times10^{\rm 6}$ to $1.101\times10^{\rm 7}$.
The NSs produced from the DD model are quite different from that of the SD model. The AIC events from the SD model may form millisecond/intermediate-mass X-ray binary pulsars and then millisecond/intermediate-mass binary pulsars that have NSs with masses of $\sim$1.25$-$1.3$\,M_{\odot}$, while the AIC events from the DD model would produce single NSs with a large mass range.
Meanwhile, AIC events from the DD model are not expected to have dense circumstellar media compared with the SD model. Thus, the strong radio emission from the shock as seen in SD model is not expected in the DD model (see Moriya 2016).
Moreover, if the formed NSs from the DD model are more massive than $\sim$$2.2\,M_{\odot}$, they will be super-massive NSs and rotation is required to support themselves (e.g. Metzger et al. 2015). In this case, the super-massive NSs may lose their rotational energy immediately via the r-mode instability and collapse into black holes. If these super-massive NSs are strongly magnetized, they may emit fast radio bursts (see Moriya 2016).

\begin{figure*}
\begin{center}
\epsfig{file=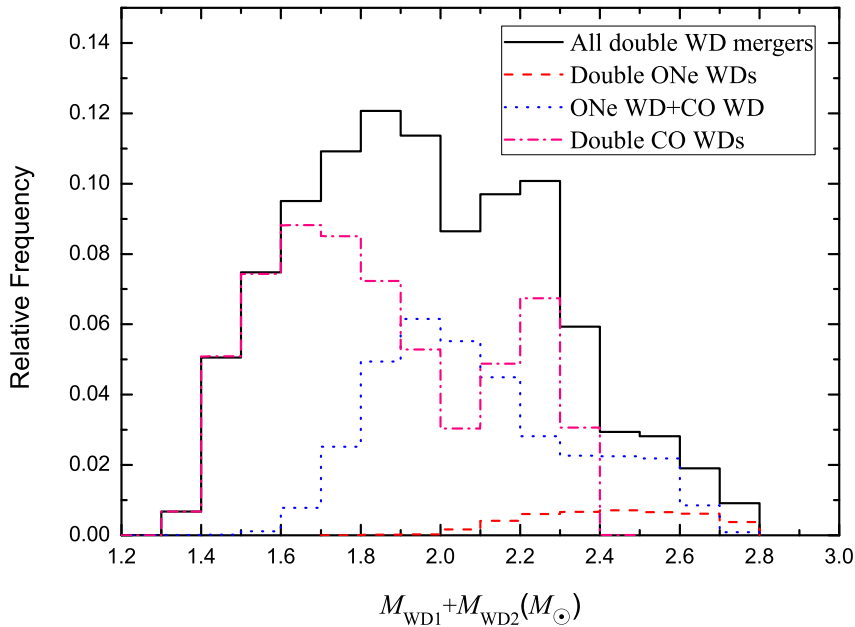,angle=0,width=14.cm}
\end{center}
\caption{The total mass distribution of the double WD mergers through AIC process with $\alpha_{\rm CE}\lambda=1.0$. The case for all double WD mergers is normalized to be 1.} 
\end{figure*}

Fig.\,6 shows the total mass distribution of double WDs that can form AIC events, which can represent the upper limit for the masses of the formed single NSs.
From this figure, we can see that the total masses of double ONe WDs mainly range from $2.0\,\rm M_{\odot}$ to twice of the Chandrasekhar limit, and the total masses of ONe WD$+$CO WD systems are mainly in the range of 1.6$-$$2.7\,\rm M_{\odot}$, and mainly range from the Chandrasekhar limit to $2.4\,\rm M_{\odot}$ for double CO WDs.
Note that the mass distribution of the single NSs from the mergers of all double WD and double CO WDs have two peaks. The larger peaks originate from the WD$+$He subgiant scenario, in which the primary WDs can grow in mass up to $0.48\,\rm M_{\odot}$ after its formation (see Sect. 2.2). For the formation of double ONe WDs and ONe WD$+$CO WD systems, the contribution of WD$+$He subgiant scenario is much smaller than that of double CO WDs. Hence, the mass distribution of single NSs from double ONe WD mergers and ONe WD$+$CO WD mergers just have single peaks.

\section{Discussions} \label{4. Discussion}
\subsection{Comparison to previous work}\label{Comparison to previous work}
Ruiter et al. (2019) recently investigated the formation of AIC events from the merging of double ONe WDs or ONe WD$+$CO WD systems in a systematic way. There are three main differences between the present work and that of Ruiter et al. (2019), as follows:
(1) We employed full stellar evolution calculations to simulate the evolution of WD$+$He star systems, while Ruiter et al. (2019) adopted analytical fitting formulae to estimate the mass accretion rate.
(2) We considered the formation of single NSs from the double CO WD merger channel, while Ruiter et al. (2019) ignored this channel. According to the present work, we found that the double CO WD merger channel may have a relatively high contribution to the formation of single NSs via AIC though there are highly uncertainties.
(3) We obtained the properties of double WDs and their gravitational wave signals (see Sect.\,4.2) from different double WD merger channels, which is not fully considered in Ruiter et al. (2019).
Note that Ruiter et al. (2019) estimated that the Galactic rate of AIC events from double WDs is $\sim$$2.3\pm0.4\times10^{\rm -4}\,\rm yr^{\rm -1}$ by using a new CE model with $\alpha_{\rm CE}=1.0$. 
The AIC rate from double WDs for the case with $\alpha_{\rm CE}\lambda=0.5$ ($\alpha_{\rm CE}=1.0$, $\lambda=0.5$) presented in this work is $\sim$$3.19\times10^{\rm -4}\,\rm yr^{\rm -1}$ (just considering the double ONe WD merger channel and the ONe WD$+$CO WD merger channel), which is roughly consistent with that of Ruiter et al. (2019).

\subsection{Gravitational wave signals}\label{Gravitational wave signals}
Recently, the ground-based aLIGO/Virgo has detected the gravitational wave signals of 10 double black hole mergers and 1 double NS merger (e.g. Abbott et al. 2016a,b, 2017a,b,c,d), which starts a new era of gravitational wave astronomy. Meanwhile, the future space-based gravitational wave detectors like LISA and TianQin would also focus on the gravitational wave signals from double WDs with short orbital periods (e.g. Marsh 2011; Luo et al. 2016).
It has been suggested that the space-based gravitational wave detectors may be capable to observe about 2700 close double WDs (Kremer et al. 2017). Here, we will provide the chirp mass distributions and gravitational wave strain amplitude of double WDs that can merge and form AIC events.

The chirp mass is a function of double WD masses that can be measured by the gravitational wave detectors directly. The chirp masses of double WDs are defined as
\begin{equation}
M_{\rm chirp}=\frac{(M_{\rm WD1}M_{\rm WD2})^{\rm 3/5}}{(M_{\rm WD1}+M_{\rm WD2})^{\rm 1/5}},
\end{equation}
where $M_{\rm WD1}$ and $M_{\rm WD2}$ are the mass of double WDs, respectively.
The gravitational wavestrain amplitude (h) represents the fractional change in separation that occurs when a gravitational wave passes through the detector. In the present work, we assume that all the binaries are in circle orbital. We adopt a dimensionless gravitational wave strain amplitude for circle orbital binaries presented by Yu \& Jeffery (2010), written as
\begin{equation}
h=5.0\times10^{\rm -22}\times(\frac{M_{\rm chirp}}{M_{\odot}})^{\rm 5/3}(\frac{P_{\rm orb}}{\rm hour})^{\rm -2/3}(\frac{d}{\rm kpc})^{\rm -1},
\end{equation}
where $P_{\rm orb}$ is the orbital period and $d$ is the distance from the double WDs to the detectors. This distance is simply assumed to be 8.5 kpc in this work.
The gravitational wave frequency of double WD binaries is defined as $f_{\rm GW}=2/P_{\rm orb}$.


\begin{figure}
\begin{center}
\begin{tabular}{@{}c@{}}
\centerline{\epsfig{file=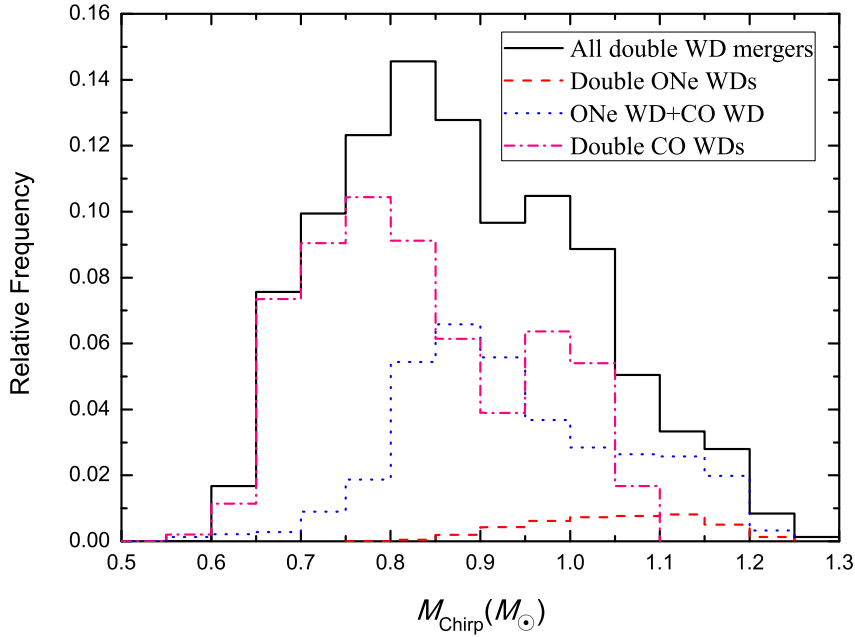,angle=0,width=14.cm}}
\end{tabular}
 \caption{Chirp mass distributions of different double WDs that can merge and produce AIC events, in which $\alpha_{\rm CE}\lambda=1.0$. The case for all double WDs here is normalized to be 1.}
\end{center}
\end{figure}

Fig.\,7 presents the chirp mass distributions of double WDs for producing AIC events at the moment of their formation.
From this figure, we can see that the chirp masses of double ONe WDs range from $0.8\, M_{\odot}$ to $1.25\, M_{\odot}$, and the distributions have a peak at $1.1\, M_{\odot}$, while the chirp masses of ONe WD$+$CO WD systems range from $0.55\, M_{\odot}$ to $1.2\, M_{\odot}$, and have a peak at $0.85\, M_{\odot}$. For the case of double CO WDs, their chirp masses distribute in the range of $0.55$$-$$1.05\, M_{\odot}$ and have two peaks at $0.75\, M_{\odot}$ and $0.95\, M_{\odot}$. The origination of these double peaks are similar to that shown in Fig.\,6.

\begin{figure}
\begin{center}
\begin{tabular}{@{}c@{}}
\centerline{\epsfig{file=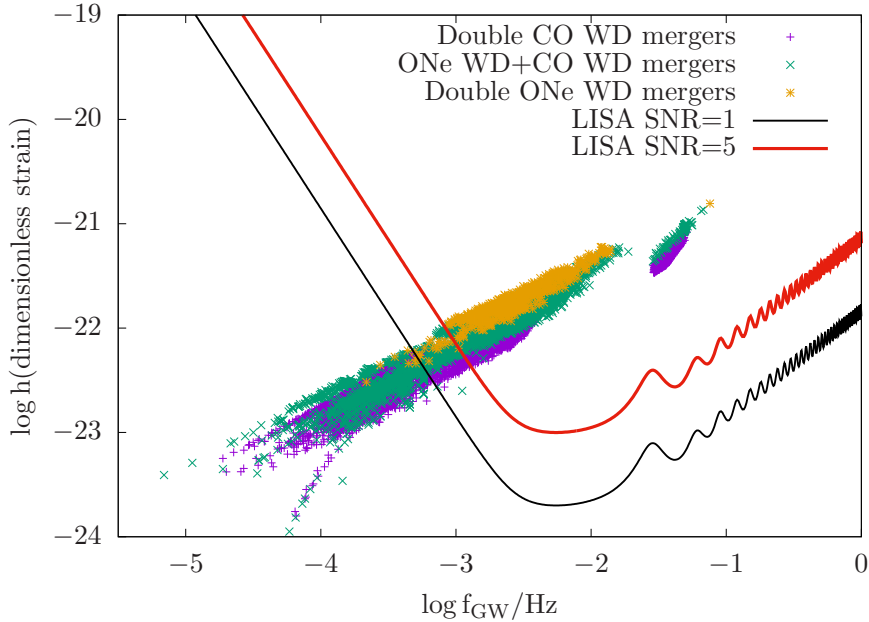,angle=0,width=14.cm}}
\end{tabular}
 \caption{Dimensionless gravitational wave strain amplitude of different double WDs that can form AIC events with $\alpha_{\rm CE}\lambda=1.0$. Here, we simply assume that the distance from double WDs to the detectors is 8.5 kpc (similar to that assumed in Ruiter et al. 2019). The sensitivity curve for the future space-based gravitational wave observatory LISA are also shown for comparison (from the online sensitivity curve generator based on Larson, Hiscock \& Hellings:
 \url{http://www.srl.caltech.edu/~shane/sensitivity/}).}
\end{center}
\end{figure}

In Fig.\,8, we present the dimensionless gravitational wave stain amplitude of double WDs that can merge and form AIC events in the $\log h-\log f_{\rm GW}$ plane. From this figure, we can see that the gravitational wave frequency of double WDs are in the range of about $10^{\rm -5}\,\rm Hz$ to $0.1\,\rm Hz$, and the dimensionless gravitational wave stain amplitude ranges from about $10^{\rm -24}$ to $10^{\rm -21}$.
The stain amplitude and frequency of double ONe WDs tend to be larger than that of ONe WD$+$CO WD sytems, and the stain amplitude and frequency of ONe WD$+$CO WD sytems is larger than that of double CO WDs.
From this figure, we can see that the distributions of double WDs here have two parts, in which the right small part originate from a specific formation scenario. In this scenario, the primordial primary fill its Roche-lobe at the thermal pulsing asymptotic giant branch star and form a CE. After the CE ejection, a WD$+$MS system will be formed. As the primordial secondary evolves, it will fill its Roche-lobe and form a CE when the secondary turns to be a red-giant or an early asymptotic giant branch star. After the ejection of the second CE, the system will evolve to a double WD system with short orbital period.
Note that the distances from the double WDs to the detectors (d) in the real Milky Way should have a wide distribution but not simply $d=8.5\,\rm kpc$, which means that the distribution of double WDs in the real Milky Way should be wider.
We also present the sensitivity curve for the future space-based gravitational wave observatory LISA with signal to noise ratio to be 1 and 5 (from the online sensitivity curve generator based on Larson, Hiscock \& Hellings
). Fig.\,8 indicates that more than half of the double WDs for producing AIC events are potential targets for the future observations of LISA.

\subsection{Detections of AIC events}\label{Detections of AIC events}
There is still no direct detection of AIC events so far. One possible reason is that the AIC events may eject quite small material (less than $0.1\,\rm M_{\odot}$), producing little $^{\rm 56}$Ni (less than $0.01\,\rm M_{\odot}$) and move at very high velocity (about 10\% of light velocity), which means that the optical transients of AIC events would be 5 magnitude (or more) fainter than a typical supernova and last for only a few days (e.g. Metzger et al. 2009; Darbha et al. 2010; Piro \& Kulkarni 2013).

Another possible reason is that it is very difficult to distinguish AIC events from other faint supernovae. The ongoing surveys may have already detected some AIC events, but they are mixed in some other transients (e.g. faint core collapse supernovae, kilonovae, Gamma-ray burst, etc) that are too difficult to distinguish them. Thus, it is very necessary to give the characteristic properties of AIC events for their identification.
Moriya (2016) investigated the observational properties of AIC events in radio frequencies, and argued that AIC events may cause fast radio bursts if a certain condition is satisfied. Moriya (2016) also suggested that gravitational waves from AIC events may be accompanied by radio-bright transients, and this can be used to confirm the AIC origin of the observed gravitational waves.
Yu, Chen \& Wang (2019) suggested that searching for dust-affected optical transients and shock-driven radio transients can help to explore the nature of super-Chandrasekhar double WD merger remnants and also the density and type ratios of double WD systems, which is beneficial in assessing their gravitational wave contributions.
Mcbrien et al. (2019) recently present the fastest declining supernova-like transient SN 2018kzr (second only to the kilonova AT2017gfo), which has the low ejecta mass, intermediate mass element composition and the high likelihood of additional powering. Mcbrien et al. (2019) argued that AIC of a WD is a possible explosion scenario for SN2018kzr.
Moriya (2019) also suggested that a recently reported radio transient in M81, VTC J095517.5$+$690813, may be caused by an AIC event of a WD.

In the present work, we suggested that there may exist $0.382\times10^{\rm 7}$ to $1.072\times10^{\rm 8}$ single NSs in the Galaxy. These single NSs may have different properties comparing with the single NSs from other formation scenarios, like their circumstellar characteristics, magnetic field, etc. Hence, we suggest that the single NSs from the AIC process originating from double WD mergers may correspond to a specific type of NSs. More theoretical and observational investigations on this type of single NSs may be quite interesting, which may also be helpful for the studies on the AIC events.


\section{Summary}\label{Summary}
In this work, we explored the properties of AIC events from the double WD mergers by considering both the contribution of the WD$+$He subgiant scenario and classical scenarios to the formation of double WDs. We conducted a large number of full stellar evolution calculations of CO/ONe WD$+$He star systems, and thus obtained the parameter space for producing AIC events via double WD mergers.
We then explored the BPS properties of AIC events and gravitational wave signals of double WDs from the WD$+$He subgiant scenario (i.e. the stable Roche-lobe overflow scenario) based on the obtained parameter space, and also considered the classical scenario for the formation of double WDs  (i.e. the CE ejection scenario).
We found that the Galactic rate of AIC events from double WD mergers is about 10\% to 2 times of the observed SN Ia rate, and the number of single NSs originating from double WD mergers range from $0.382\times10^{\rm 7}$ to $1.072\times10^{\rm 8}$ for different models in our Galaxy.
This work also indicates that the double WD chirp masses range from $0.55\, M_{\odot}$ to $1.25\, M_{\odot}$, and more than half of the double WDs for producing AIC events are potential objects for the future observations of LISA.
However, it still needs more numerical simulations on the AIC process of double WD mergers and more observational surveys to identify AIC events.

\section*{Acknowledgments}
We acknowledge useful comments and suggestions from the anonymous referee.
We thank Ashley Ruiter for her kind comments and helpful discussions.
The present work is supported by the Natural Science Foundation of China (Nos 11903075, 11873085, 11673059 and
11521303), Chinese Academy of Sciences (No. QYZDB-SSW-SYS001), the Western Light Youth Project of Chinese Academy of Sciences and the Yunnan Province (Nos 2018FB005 and 2019FJ001).

\label{lastpage}
\end{document}